# Determination of thermal emission spectra maximizing thermophotovoltaic performance using a genetic algorithm

John DeSutter[†], Michael P. Bernardi and Mathieu Francoeur[†]

*Radiative Energy Transfer Lab, Department of Mechanical Engineering, University of Utah, Salt Lake City, Utah 84112, USA*

**Abstract**

Optimal radiator thermal emission spectra maximizing thermophotovoltaic (TPV) conversion efficiency and output power density are determined when temperature effects in the cell are considered. To do this, a framework is designed in which a TPV model that accounts for radiative, electrical and thermal losses is coupled with a genetic algorithm. The TPV device under study involves a spectrally selective radiator at a temperature of 2000 K, a gallium antimonide cell, and a cell thermal management system characterized by a fluid temperature and a heat transfer coefficient of 293 K and 600 $Wm^{-2}K^{-1}$. It is shown that a maximum conversion efficiency of 38.8% is achievable with an emission spectrum that has emissivity of unity between 0.719 eV and 0.763 eV and zero elsewhere. This optimal spectrum is less than half of the width of those when thermal losses are neglected. A maximum output power density of 41708 $Wm^{-2}$ is achievable with a spectrum having emissivity values of unity between 0.684 eV and 1.082 eV and zero elsewhere when thermal losses are accounted for. These emission spectra are shown to greatly outperform blackbody and tungsten radiators, and could be obtained using artificial structures such as metamaterials or photonic crystals.

---

[†] Corresponding authors. Tel.: +1 801 581 5721, Fax: +1 801 585 9825

E-mail addresses: mfrancoeur@mech.utah.edu (M. Francoeur), john.desutter@utah.edu (J. DeSutter)

## 1. Introduction

Thermophotovoltaic (TPV) power generators are conceptually similar to solar photovoltaic (PV) cells in which thermal radiation is directly converted into electricity. Solar PV cells are irradiated by the sun while TPV power generators have a radiator component, in addition to the cell, that is heated by an external source [1,2]. The radiator and the cell are separated by a vacuum gap, and the external source of energy can potentially be anything that produces heat. Therefore, TPV power generation is a promising technology for the cogeneration of heat and electricity in residential appliances [3] and for recycling wasted heat in engines and industrial production processes [1] to name only a few. An overall TPV efficiency (heat source to electricity) of 2.5% has been experimentally demonstrated when heating a selectively emitting one-dimensional photonic crystal radiator using a combustion process [4]. TPV devices can also be used to harness solar energy in which the radiator is an intermediate layer between the sun and the cell. The intermediate layer absorbs solar radiation and reradiates this energy towards the cell. This allows for the radiation to be spectrally emitted and/or filtered in a way that better matches the absorption characteristics of the cell. It has been shown that the theoretical maximum overall efficiency (sun to electricity) for a solar TPV power generator that radiates monochromatically at a frequency corresponding to the absorption bandgap of the cell is 85.4% [5]. This efficiency assumes idealities that are likely not achievable in practice such as an infinite emitter area and monochromatic emission from the radiator. Other work has been conducted on global optimization of TPV devices to improve overall efficiency for more realistic cases consisting of a finite emitter area and broadband radiator emission [6-9]. The current experimental record for overall solar TPV efficiency is 3.2% [10]. Despite this important milestone, solar PV cells still largely outperform solar TPV power generators.

The low experimental overall efficiency is mostly due to low TPV conversion efficiency (radiator to electricity) [4,10,11]. In particular, Datas and Algora [11] pointed out that the low TPV conversion efficiency is largely caused by thermal losses leading to an overheating of the cell. Thermal losses are due to absorption of radiation by the free carriers and the lattice, non-radiative recombination of electron-hole pairs (EHPs) and thermalization of radiation with energy larger than the bandgap. Thermal losses negatively affect TPV performance by increasing the dark current, due to an increase of the cell temperature, which opposes the generated photocurrent [12,13]. By taking into account thermal losses in a TPV system capitalizing on the near-field effects of thermal radiation, it was shown in Ref. [13] that there is a high-energy cutoff in the radiator emission spectrum above which radiation has a net negative effect on TPV output power density. In addition to thermal losses, radiative and electrical losses in the cell also negatively affect TPV performance. Radiation absorbed by the cell with energy smaller than the bandgap that does not contribute to photocurrent generation is a radiative loss. Electrical losses are caused by EHPs recombining before reaching the depletion region, thus not contributing to photocurrent generation. Many papers have been devoted to the design of selectively emitting radiators in an effort to maximize TPV performance when accounting for only radiative losses [e.g. 14,15], and radiative and electrical losses [e.g. 7,16-21]. An optimal radiator design, however, must also account for thermal losses in the cell due to their large impact on TPV performance. To the best of our knowledge, no attempt has been made to design radiator emission spectra maximizing TPV performance while taking into account all three loss mechanisms.

The objective of this work is to determine radiator emission spectra maximizing TPV conversion efficiency and output power density while accounting for all loss mechanisms in the cell. This is achieved via a rigorous optimization framework in which a multi-physics model combining radiative, electrical and thermal transport [12] is coupled with the publicly available genetic algorithm (GA) PIKAIA [22]. The TPV power generator analyzed hereafter consists of a spectrally selective radiator, a cell made of gallium antimonide (GaSb) and a cell thermal management system. The TPV-GA framework is described in section 2. In section 3, radiator emission spectra maximizing TPV conversion efficiency and output power density are discussed, and TPV performances with the optimal spectra are compared against those obtained with blackbody and tungsten radiators. Concluding remarks are provided in section 4.

## 2. Description of the thermophotovoltaic (TPV)-genetic algorithm (GA) model

When radiative, electrical and thermal losses are taken into account, determining the radiator emission spectrum maximizing TPV performance (conversion efficiency $\eta$ or output power density $P_m$) is not a straightforward task, since these three loss mechanisms are strongly coupled with each other [13]. As such, it is very difficult, if not impossible, to derive a closed-form solution providing optimal emission spectrum. To find the emission spectrum maximizing TPV performance, a multi-physics model combining radiative, electrical and thermal transport in TPV devices [12] is coupled with a GA [22].

### 2.1. TPV model

The TPV power generator analyzed in this work, shown in Fig. 1, is modeled as a one-dimensional system for which only the variations along the $z$-direction are taken into account. The spectrally selective radiator, denoted as layer 0, is a semi-infinite medium maintained at a constant and uniform temperature $T_{rad}$. The cell consists of a p-n junction of GaSb, where layers 2 and 3 are respectively the p-doped (thickness of 0.4 µm, doping concentration of $10^{19}$ cm$^{-3}$) and n-doped (thickness of 10 µm, doping concentration of $10^{17}$ cm$^{-3}$) regions. The absorption bandgap $E_g$ of the GaSb cell at a temperature of 293 K is 0.723 eV (angular frequency $\omega_g$ of $1.10 \times 10^{15}$ rad/s; wavelength $\lambda_g$ of 1.71 µm) [12]. Note that GaSb is chosen as the cell material since its absorption bandgap in the near infrared matches the dominant wavelength emitted by typical TPV radiators ($T_{rad} \sim$ 1300-2000 K) [23]. The radiator and the cell are separated by a vacuum gap of thickness $d$ denoted as layer 1. Since all layers are assumed to be infinite along the $\rho$-direction, the view factor between the radiator and the cell is unity. In addition, the vacuum gap thickness $d$ is assumed to be much larger than the dominant wavelength emitted, such that radiation heat transfer occurs exclusively via propagating modes [24]. Layer 4 is the cell thermal management system described by a fluid temperature $T_\infty$ and a heat transfer coefficient $h_\infty$.

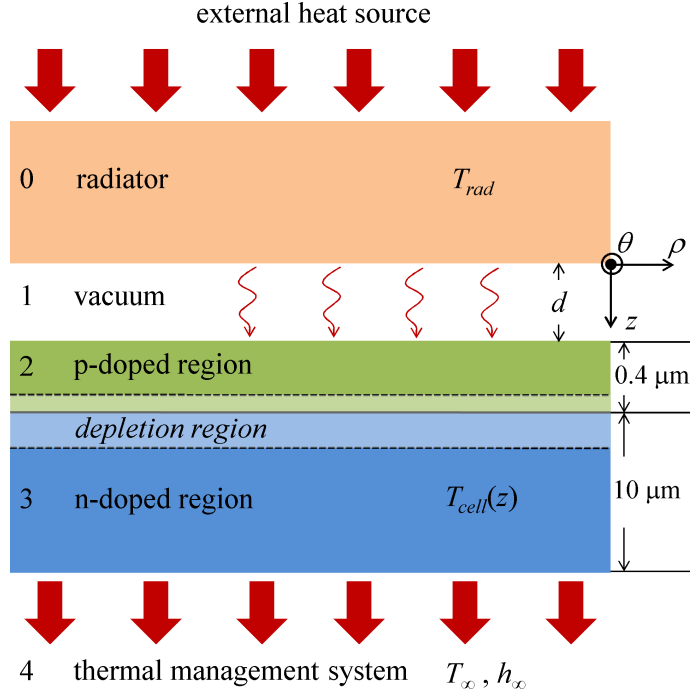

Figure 1. Schematic representation of the TPV power generator under study: A spectrally selective radiator at temperature $T_{rad}$ is separated from a GaSb cell by a vacuum gap of thickness $d$ that is much larger than the dominant wavelength emitted.

TPV simulations are performed by discretizing the cell into $N$ control volumes of thickness $\Delta z_j$. Radiation transport is modeled using fluctuational electrodynamics [25] in which stochastic currents are added into Maxwell's equations to account for thermal emission. This formalism has the advantage of being valid both in the far- and near-field regimes of thermal radiation [24]. The monochromatic radiative heat flux absorbed by a control volume $\Delta z_j$ due to thermal emission by the radiator is given by [26]:

$$q_{\omega,abs}^{\Delta z_j} = \Theta(\omega, T_{rad}) \int_0^\infty \frac{k_\rho dk_\rho}{4\pi^2} \left[ \overline{T}(k_\rho, z_j, \omega) - \overline{T}(k_\rho, z_{j+1}, \omega) \right] \qquad (1)$$

where $\Theta$ is the mean energy of a Planck oscillator, $k_\rho$ is the wavevector component along the $\rho$-direction, $\overline{T}$ is the energy transmission factor while $z_j$ and $z_{j+1}$ are the boundaries delimiting the control volume $\Delta z_j$. The energy transmission factors in Eq. (1) are calculated via dyadic Green's functions for layered media; the computational details can be found in Ref. [27]. Note that the radiative flux absorbed by the radiator due to thermal emission by the control volume $\Delta z_j$ is calculated with Eq. (1) using $\Theta(\omega,T_{cell,j})$, instead of $\Theta(\omega,T_{rad})$, where $T_{cell,j}$ is the temperature of the cell within $\Delta z_j$.

Once Eq. (1) is solved, the local monochromatic generation rate of EHPs $g_\omega(z)$ is calculated using the radiation absorbed by the cell [12]. EHPs are generated only when the radiation energy is equal to or larger than the cell absorption bandgap. Radiation transport is then coupled with electrical transport by adding $g_\omega(z)$ to the minority carrier diffusion equations [12]:

$$D_{(e,h)} \frac{d^2 \Delta n_{(e,h),\omega}(z)}{dz^2} - \frac{\Delta n_{(e,h),\omega}(z)}{\tau_{(e,h)}} + g_\omega(z) = 0 \qquad (2)$$

where the subscripts $e$ and $h$ denote electrons, the minority carriers in the p-doped region, and holes, the minority carriers in the n-doped region. The variable $D_{(e,h)}$ is the minority carrier diffusion coefficient, $\Delta n_{(e,h),\omega}$ is the local excess of minority carriers above equilibrium concentration, while $\tau_{(e,h)}$ is the minority carrier lifetime that accounts for radiative, Auger and Shockley-Read-Hall recombination. Surface recombination of EHPs is included in the boundary conditions at the vacuum-cell and cell-thermal management system interfaces; at the boundaries of the depletion region, it is assumed that minority carriers are swept by the electric field of the p-n junction [12]. The monochromatic photocurrent generated within the depletion region is calculated directly using $g_\omega(z)$, while the monochromatic photocurrent generated at the boundaries of the depletion region is determined from $\Delta n_{(e,h),\omega}$ obtained by solving Eq. (2) [12].

The total photocurrent $J_{ph}$ is afterwards computed by integrating the sum of the aforementioned contributions from $\omega_g$ to infinity.

When thermal losses are included in the model, the solution of the minority carrier diffusion equations is also used to compute heat generation in the cell by bulk non-radiative recombination $Q_{NRR}$ and radiative recombination $Q_{RR}$ of EHPs. In addition, the local radiative source term $Q_{LFC}$, representing the balance between thermal emission and absorption by the lattice and the free carriers, and heat dissipation by thermalization $Q_T$ are calculated from the solution of the radiation transport equation. These four contributions are included in the one-dimensional steady-state energy equation to compute the cell temperature distribution:

$$k\frac{d^2 T_{cell}(z)}{dz^2} + Q_T(z) + Q_{NRR}(z) - Q_{LFC}(z) - Q_{RR}(z) = 0 \tag{3}$$

where $k$ is the thermal conductivity of the cell. Internal heat conduction and surface recombination of EHPs at the boundaries of the cell are balanced with external heat fluxes. The external heat flux is zero at the vacuum-cell interface, while an external heat flux due to convection with $T_\infty$ and $h_\infty$ is imposed at the cell-thermal management system interface. After solving Eq. (3), an updated temperature distribution within the cell, $T_{cell}(z)$, is obtained. The radiative, electrical and thermophysical properties of the cell, given in Ref. [12], are computed at the updated temperature, and calculations are repeated until $T_{cell}(z)$ converges. In post-processing, the dark current density $J_0(V)$ is determined by solving Eq. (2) without the generation rate of EHPs for a series of voltage $V$, and the current density generated is computed as $J(V) = J_{ph} - J_0(V)$. Note that the contributions of the bias voltage and temperature to the dark current density are included in the boundary conditions imposed at the edges of the depletion region [12]. The maximum output power density $P_m$ is determined directly from the current-voltage ($J$-$V$) characteristic of the cell [12], while the conversion efficiency is calculated as follows:

$$\eta = \frac{P_m}{q_{abs}^{cell} + q_{trans}^{cell}} \tag{4}$$

where $q_{abs}^{cell}$ and $q_{trans}^{cell}$ are respectively the total radiative heat flux absorbed and transmitted by the cell. Note that since the view factor between the radiator and the cell is unity, the radiation reflected by the cell is not considered as a loss.

## 2.2. Coupling of the TPV model with the GA

To determine the radiator emission spectrum maximizing TPV performance ($P_m$ or $\eta$), the publicly available GA PIKAIA [22], which is an optimization tool utilizing the concept of evolution, is coupled with the TPV model. For this purpose, the radiator emission spectrum is discretized into $M$ frequency bands and is characterized by $M$ values of spectral, hemispherical emissivity (hemispherical will be omitted hereafter). The GA operates with the goal of finding the individual, or emission spectrum in the case of this paper, consisting of a set of traits, or spectral emissivity values, maximizing a user defined objective function which is either conversion efficiency $\eta$ or power density $P_m$ in this work. GA computations are initialized by generating a random population of emission spectra. For each emission spectrum in the population, $P_m$ or $\eta$ is calculated using the TPV model. The GA then ranks each emission spectrum based on $P_m$ or $\eta$. Spectra in the population are selected as parents using the roulette wheel algorithm in which the share of the wheel is determined by rank. For instance, the highest ranked emission spectrum has the largest share of the wheel, and thus has the highest probability of being selected as a parent, while the lowest ranked spectrum has the smallest share of the wheel. The parent emission spectra breed to create offspring emission spectra consisting of a combination of spectral emissivity values from both parents. To avoid convergence at false maxima, mutations involving a small change to a single spectral emissivity value of an offspring are introduced. Some offspring may not experience any mutations while others may experience several. Offspring then randomly replace emission spectra in the parent generation. However, the emission spectrum with the highest ranking in the parent generation can only be replaced by an offspring producing a superior $P_m$ or $\eta$. This ensures that the best emission spectrum is not discarded. The population is evolved in this manner over a user specified number of generations in order to determine an emission spectrum maximizing $P_m$ or $\eta$.

Following the optimization procedure described above, the output power density $P_m$ or conversion efficiency $\eta$ is calculated approximately two hundred thousand times with the TPV model. Since the TPV model requires approximately five minutes on a personal desktop computer (Intel® Core ™ i7-4790 processor: 4 cores clocked at 3.6 GHz, 16 GB Memory), mostly due to calculating radiative transport, one evolutionary period would take nearly two years to complete. To circumvent this problem and to avoid using a large computer cluster, a database has been created in which the spectral radiative heat flux absorbed within each control volume has been recorded for a fixed blackbody radiator temperature and for various cell temperatures. As the radiator is assumed to be a blackbody, the actual radiative heat flux absorbed is readily obtained by multiplying the spectral flux calculated from Eq. (1) by the corresponding spectral emissivity. This approach reduces total computational time to approximately fifteen hours.

The TPV-GA algorithm for determining a radiator emission spectrum maximizing power density $P_m$ or conversion efficiency $\eta$ is summarized as follows:

1. For a fixed radiator temperature $T_{rad}$, specify the spectral limits and the spectral discretization (i.e., number of spectral bands $M$).

2. Using Eq. (1), generate a database of radiative heat flux values for a blackbody radiator at temperature $T_{rad}$ for various cell temperatures $T_{cell}$. For a given $T_{rad}$ and $T_{cell}$, the database provides the radiative heat flux absorbed within a given control volume $\Delta z_j$ for each spectral band.

3. Run the GA:

   i) An initial random population of emission spectra is generated.

   ii) For each emission spectrum in the population, the power density $P_m$ or conversion efficiency $\eta$ is obtained via the TPV model. The emission spectra are then ranked based on maximizing $P_m$ or $\eta$.

   iii) Emission spectra within the population are selected as parents based on rank.

   iv) Offspring emission spectra are generated by breeding parent emission spectra.

   v) Mutations are performed on offspring emission spectra.

   vi) Emission spectra from the parent generation are randomly replaced by offspring emission spectra.

   vii) Steps ii) to vi) are repeated until a user defined evolutionary period is completed.

   viii) The emission spectrum producing the highest power density $P_m$ or conversion efficiency $\eta$ is selected.

4. The emission spectrum selected in step 3 is verified by performing direct calculation of $P_m$ or $\eta$ using the forward TPV model in which heat flux values are calculated.

## 3. Results

In this section, radiator emission spectra maximizing TPV conversion efficiency $\eta$ and output power density $P_m$ obtained from the TPV-GA model are analyzed. The radiator temperature $T_{rad}$ is fixed at 2000 K, while the thermal management system is characterized by a fluid temperature $T_\infty$ and a heat transfer coefficient $h_\infty$ of 293 K and 600 Wm$^{-2}$K$^{-1}$, respectively. A heat transfer coefficient of 600 Wm$^{-2}$K$^{-1}$ is achievable via forced convection with a liquid. Note that below a $h_\infty$ value of 600 Wm$^{-2}$K$^{-1}$, the cell temperature exceeds its melting point when a blackbody radiator at 2000 K is used. The p-doped and n-doped regions are respectively discretized into 400 and 800 control volumes. The thickness of the depletion region, assumed to be exclusively in the n-doped region, is temperature-dependent [12]; at a temperature of 293 K, the depletion region is 113-nm-thick. In all simulations, a convergence criterion of $10^{-4}$ on the cell temperature is used. It is also worth noting that the temperature gradient in the cell is negligible (maximum temperature difference less than 0.1 K), such that an averaged cell temperature $T_{cell}$ is reported.

Additionally, for all simulations, a population size of 100 with an evolutionary period of 2000 generations are sufficient for determining emission spectra maximizing $P_m$ or $\eta$. The cell temperature discretizations for heat flux values recorded in the database are 10 K and 1 K when maximizing $P_m$ and $\eta$, respectively. The coarser discretization employed when maximizing $P_m$ is due to the fact that this case leads to larger temperature fluctuations than when $\eta$ is maximized. Refining the temperature discretization did not change the emission spectrum maximizing $P_m$ determined with the GA. Note that linear interpolation is used to retrieve radiative heat fluxes in the database.

### 3.1. Maximization of TPV conversion efficiency

The emission spectrum maximizing conversion efficiency $\eta$ when considering radiative, electrical and thermal losses (RET) is shown in Fig. 2. This spectrum has been obtained from the TPV-GA framework using 125 spectral bands, each having a bandwidth of $1.42 \times 10^{-3}$ eV, within the range of 0.711 eV to 0.889 eV in an effort to minimize computation time. Indeed, a pre-analysis utilizing a larger bandwidth covering the entire radiator emission spectrum revealed that the emissivity outside the aforementioned thresholds must be zero. Note that further refinement of the spectral discretization from 125 to 250 bands resulted in variations of the conversion efficiency, the power density and the width and spectral cutoffs of the optimal emission spectrum of less than 1%.

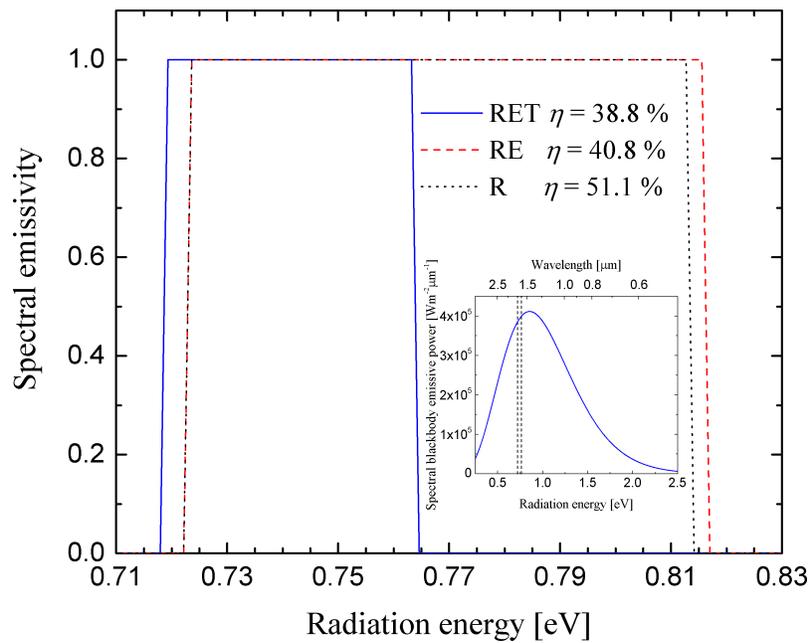

Figure 2. Thermal emission spectra maximizing conversion efficiency $\eta$ when radiative (R), radiative and electrical (RE), and radiative, electrical and thermal losses (RET) are considered. The inset shows the portion of the blackbody emissive power at 2000 K included in the RET optimal spectrum.

The emission spectrum maximizing $\eta$ is not simply monochromatic at the cell absorption bandgap $E_g$, but takes the form of a narrow step function, where the emissivity below 0.719 eV (low-energy cutoff $E_{lc}$) and above 0.763 eV (high-energy cutoff $E_{hc}$) is zero, while the emissivity between these limits has the maximum value of unity. This optimal emission spectrum results in a conversion efficiency $\eta$ of 38.8%, a power density $P_m$ of 10101 Wm$^{-2}$ and a cell temperature $T_{cell}$ of 302 K. The low-energy cutoff corresponds to the absorption bandgap of GaSb at a temperature of 302 K; the spectral location of the high-energy cutoff will be explained later when analyzing Fig. 4. For comparison, the emission spectra maximizing $\eta$ when considering only radiative losses (R) and radiative and electrical losses (RE) are also shown in Fig. 2. For the R case, it is assumed that all EHPs generated contribute to photocurrent, but Auger, Shockley-Read-Hall and radiative recombination are still accounted for when calculating the dark current [28]. In both the R and RE cases, the cell temperature is fixed at 293 K. The optimal emission spectrum for the RET case has less than half of the width of the spectra obtained for the R and RE cases. The low- and high-energy cutoffs, conversion efficiency and power density for the R case are respectively 0.723 eV (cell absorption bandgap at 293 K), 0.813 eV, 51.1% and 25276 Wm$^{-2}$; the corresponding values for the RE case are 0.723 eV, 0.816 eV, 40.8% and 20793 Wm$^{-2}$. It should be noted that while the optimal emission spectra are indeed step functions consisting solely of the maximum and minimum possible values of emissivity, the optimization analysis accounted for all possible emissivity values between 0 and 1.

The $J$-$V$ characteristics of the cell for the three emission spectra shown in Fig. 2 are presented in Fig. 3. The $J$-$V$ characteristic obtained from a quasi-monochromatic radiator when all losses are taken into account is also displayed in Fig. 3. Note that the quasi-monochromatic emission spectrum has low- and high-energy cutoffs of 0.723 eV and 0.725 eV; within these spectral limits, the radiator has an emissivity of unity. The conversion efficiency, power density and cell temperature for this case are respectively 31.7%, 267 Wm$^{-2}$ and 293.3 K.

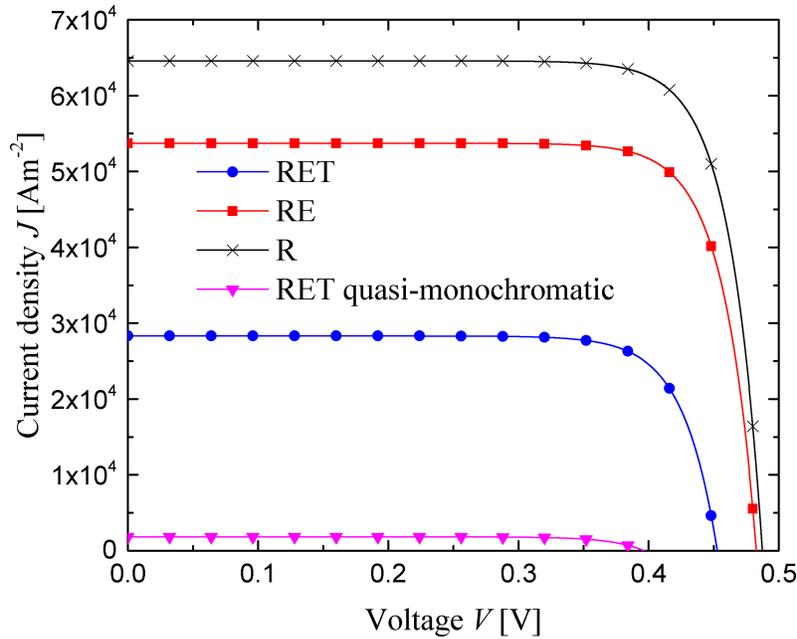

Figure 3. *J-V* characteristics obtained from the emission spectra maximizing conversion efficiency $\eta$ when radiative (R), radiative and electrical (RE), and radiative, electrical and thermal losses (RET) are considered. The *J-V* characteristic obtained from a quasi-monochromatic radiator when considering all loss mechanisms (RET) is also shown for comparison.

The short-circuit current density $J_{sc}$ (i.e., current density when $V = 0$) is significantly larger for all optimal emission spectra compared to the quasi-monochromatic case. This is due to a larger amount of radiation absorbed by the cell with energy greater than its bandgap, thus leading to a larger photocurrent generation. As expected, electrical losses cause the short-circuit current density to decrease. The decrease in $J_{sc}$ from the RE to the RET case is attributed to thermal losses in addition to a smaller amount of radiation absorbed by the cell with energy higher than its bandgap. Furthermore, for a fixed cell temperature, the open-circuit voltage $V_{oc}$ (i.e., voltage when $J = 0$) increases with increasing photocurrent generation. For instance, for the R and RE cases where the cell temperature is fixed at 293 K, the dark current density $J_0$ as a function of the voltage $V$ is the same, but the open-circuit voltage for the R case is slightly higher due to a larger photocurrent generation. When all losses are considered, the open-circuit voltage is smaller than for the R and RE cases due to a lower photocurrent generation and a larger cell temperature ($J_0$ increases as $T_{cell}$ increases [12]).

The $J$-$V$ characteristics in Fig. 3 clearly demonstrate that a quasi-monochromatic radiator leads to a low power density. However, Fig. 3 gives little physical insight regarding the high-energy cutoff $E_{hc}$ of the optimal emission spectra, and does not explain why the emission spectrum maximizing $\eta$ is not simply monochromatic near the cell absorption bandgap. These can be explained by writing the conversion efficiency in terms of the loss mechanisms. The output power density $P_m$ in Eq. (4) can be written as $J_m V_m$, where $V_m$ is the voltage at maximum power and $J_m$ ($= J(V_m)$) is the current density at maximum power. The current density $J_m$ can also be expressed in terms of photocurrent density $J_{ph}$ and dark current density $J_0(V_m)$ as follows:

$$J_m = J_{ph} - J_0(V_m) \tag{5}$$

The photocurrent density $J_{ph}$ can be written in terms of the radiative flux absorbed by the cell $q_{abs}^{cell}$, the energy flux lost to thermalization in the cell $q_T^{cell}$ and the energy flux lost to EHP recombination in the cell $q_R^{cell}$ (includes bulk radiative and non-radiative recombination as well as surface recombination):

$$J_{ph} = \left( \frac{q_{abs}^{cell} - q_T^{cell}}{E_g} \right) - \left( \frac{q_R^{cell}}{E_g} \right) \tag{6}$$

where the fluxes due to thermalization and bulk recombination have been obtained by integrating the volumetric heat sources $Q_T$, $Q_{RR}$ and $Q_{NRR}$, given in Eq. (3), over the thickness of the cell. Equation (6) is only valid when all radiation absorbed by the cell has energy equal to or larger than $E_g$, as is the case for the optimal emission spectra of Fig. 2. In addition, it is implicitly assumed in Eq. (6) that absorption by the lattice and the free carriers is negligible when the radiation energy $E$ is equal to or larger than $E_g$. This approximation is acceptable, as the contribution from the lattice and the free carriers in GaSb for $E \geq E_g$ is much smaller than interband absorption [12]. The first term on the right-hand side of Eq. (6) is interpreted as a current density due to EHP generation ($J_G^{EHP}$), while the second term is a current density due to EHP recombination ($J_R^{EHP}$). Substitution of Eqs. (5) and (6) into Eq. (4) leads to the following expression for the conversion efficiency:

$$\eta = \frac{[J_G^{EHP} - J_R^{EHP} - J_0(V_m)]}{q_{abs}^{cell} + q_{trans}^{cell}} V_m \equiv [\tilde{J}_G^{EHP} - \tilde{J}_R^{EHP} - \tilde{J}_0(V_m)]V_m \equiv \tilde{J}_m V_m \tag{7}$$

where ~ indicates effective current density (i.e., current density divided by the sum of the radiative flux absorbed and transmitted by the cell ($q_{abs}^{cell} + q_{trans}^{cell}$). Note that all terms in Eq. (7) depend on the cell temperature. The effective current density due to EHP generation $\tilde{J}_G^{EHP}$, the effective current density due to EHP recombination $\tilde{J}_R^{EHP}$, and the effective dark current density at maximum power $\tilde{J}_0(V_m)$ as a function of the high-energy cutoff $E_{hc}$ are shown in Fig. 4(a); the effective current density at maximum power $\tilde{J}_m$, the voltage at maximum power $V_m$, and the conversion efficiency $\eta$ as a function of $E_{hc}$ are presented in Fig. 4(b). Note that for the purpose of comparison, the low-energy cutoff is equal to the cell absorption bandgap at a temperature of 293 K for the R, RE and RET cases.

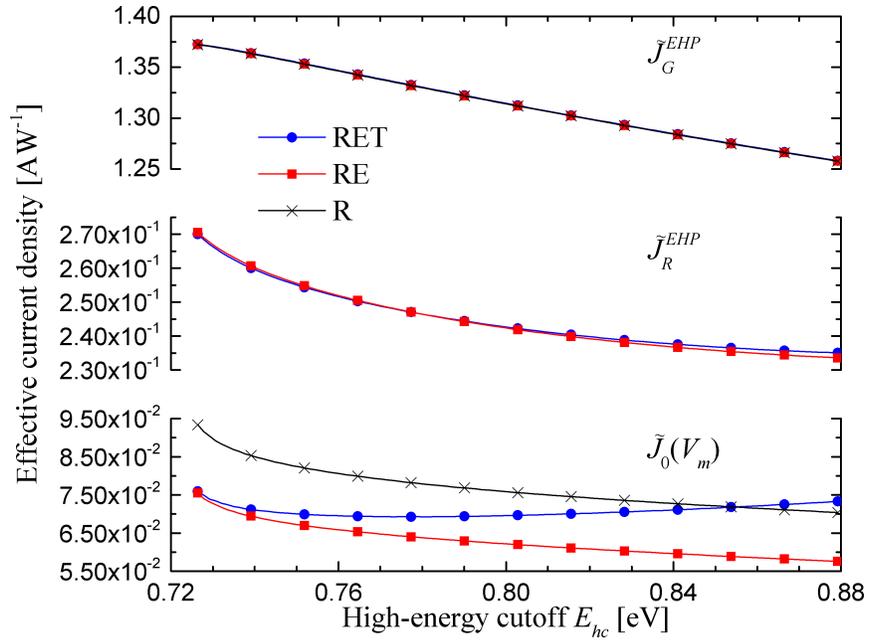

(a)

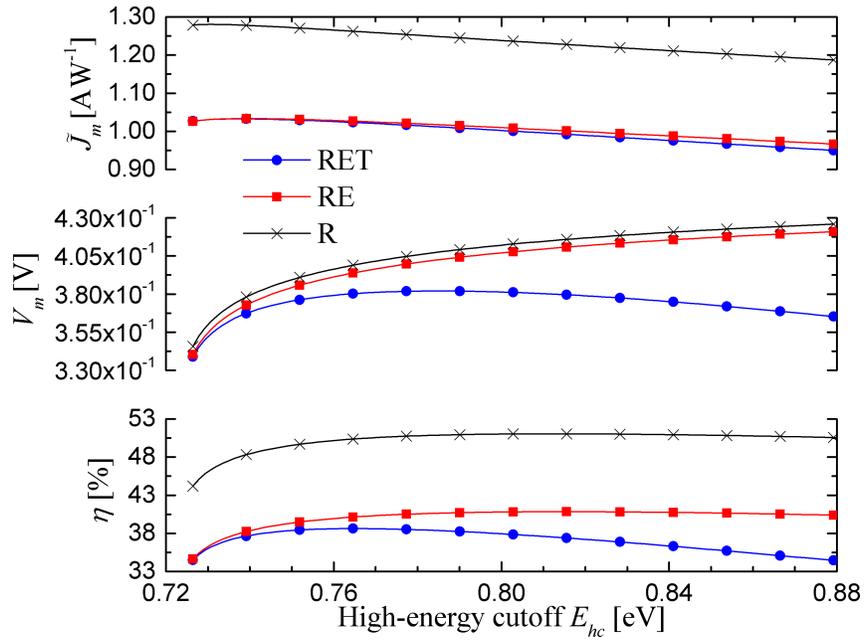

(b)

Figure 4. (a) Effective current density due to EHP generation $\tilde{J}_G^{EHP}$, effective current density due to EHP recombination $\tilde{J}_R^{EHP}$, and effective dark current density at maximum power $\tilde{J}_0(V_m)$ as a function of the high-energy cutoff $E_{hc}$ for the RET, RE and R cases. (b) Effective current density at maximum power $\tilde{J}_m$, voltage at maximum power $V_m$, and conversion efficiency $\eta$ as a function of the high-energy cutoff $E_{hc}$ for the RET, RE and R cases.

It can be seen in Fig. 4(a) that for all cases, $\tilde{J}_G^{EHP}$ decreases as the high-energy cutoff $E_{hc}$ increases due to increasing loss by thermalization (i.e., a smaller portion of the radiation energy contributes to EHP generation as $E_{hc}$ increases). Figure 4(a) also shows that $\tilde{J}_R^{EHP}$ is slightly different for the RE and RET cases since the electrical properties of the cell are temperature-dependent, thus affecting the electrical losses [12,13]. When accounting for thermal losses, the cell temperature increases with $E_{hc}$ due to increasing heat dissipation by thermalization and EHP recombination. The rapid decrease in $\tilde{J}_R^{EHP}$ at low $E_{hc}$ is caused by a decreasing $\tilde{J}_G^{EHP}$ in addition to a decreasing proportion of EHPs that recombine before reaching the depletion region. Indeed, when $E_{hc}$ is 0.726 eV, 19.8% of EHPs are generated inside and within 200 nm above and below the depletion region compared to 24.8% when $E_{hc}$ is 0.780 eV for the RE case. Note that $\tilde{J}_R^{EHP}$ is zero when only radiative losses are considered. When thermal losses are neglected, $\tilde{J}_0(V_m)$ decreases as the high-energy cutoff increases. The actual dark current density increases with $E_{hc}$, but Fig. 4(a) shows that its contribution to the conversion efficiency decreases with $E_{hc}$. On the other hand, when thermal losses are taken into account, $\tilde{J}_0(V_m)$ increases starting at an $E_{hc}$ value of 0.777 eV due to an increasing cell temperature. This shows that the actual dark current density $J_0(V_m)$ has a larger impact on TPV performance when thermal losses are accounted for. The dark current density causes the voltage at maximum power $V_m$ to decrease starting at an $E_{hc}$ value of 0.784 eV for the RET case (Fig. 4(b)). The top panel of Fig. 4(b) shows that $\tilde{J}_m$ slightly increases for low values of $E_{hc}$ near the cell absorption bandgap, and afterwards decreases with increasing high-energy cutoff, regardless of the losses considered. While $\tilde{J}_G^{EHP}$ indeed decreases with $E_{hc}$, the large drop of $\tilde{J}_R^{EHP}$ and $\tilde{J}_0(V_m)$ at small $E_{hc}$ values cause $\tilde{J}_m$ to initially increase.

It is clear from Fig. 4(b) that a quasi-monochromatic radiator with emission near the absorption bandgap of the cell does not lead to maximum efficiency. Indeed, for small $E_{hc}$ values, $\tilde{J}_m$ first increases before starting to decrease with increasing $E_{hc}$. When all losses are taken into account, the increasing rate of $V_m$ outweighs the decreasing rate of $\tilde{J}_m$ until an $E_{hc}$ value of 0.765 eV (when $E_{lc}$ equals $E_g$ at 293 K). For this case, the spectral location of the high-energy cutoff is largely dictated by the cell temperature which negatively affects both voltage at maximum power $V_m$ and effective dark current density $\tilde{J}_0(V_m)$. When thermal losses are neglected, $V_m$ and $\tilde{J}_0(V_m)$ do not exhibit inflexion points ($T_{cell}$ is constant) such that the $E_{hc}$ value leading to maximum conversion efficiency is significantly higher for the R and RE cases.

As observed in Fig. 2, the high-energy cutoff $E_{hc}$ of the emission spectrum maximizing conversion efficiency is larger in the RE case when compared to the R case. This behavior is caused by the effective current density due to EHP recombination $\tilde{J}_R^{EHP}$. Indeed, for both the R and RE cases, the increasing rate of $V_m$ is essentially the same (Fig. 4(b)). However, when electrical losses are accounted for, the rapid decrease of $\tilde{J}_R^{EHP}$ slows down the drop of $\tilde{J}_m$ as a function of $E_{hc}$ when compared to the R case for which $\tilde{J}_R^{EHP}$ is zero. A slower decrease in $\tilde{J}_m$ thus pushes the high-energy cutoff of the optimal emission spectrum toward a slightly larger value.

It is also interesting to analyze the sensitivity of the conversion efficiency to the low-energy cutoff $E_{lc}$ of the optimal emission spectrum when considering all loss mechanisms. If $E_{lc}$ decreases slightly below the absorption bandgap of the cell $E_g$, the conversion efficiency dramatically decreases since none of the radiation with energy lower than $E_g$ contributes to the output power density. For instance, when the low-energy cutoff $E_{lc}$ is 0.01 eV and 0.02 eV below $E_g$, the conversion efficiency, which was originally 38.8%, decreases to 32.9% and 28.5%. Conversely, an $E_{lc}$ value slightly higher than $E_g$ does not have a large impact on the conversion efficiency. If the low-energy cutoff is 0.01 eV and 0.02 eV above $E_g$, the resulting conversion efficiencies are respectively 38.3% and 37.6%. These slightly lower values of $\eta$ are attributed to the decreasing amount of radiation contributing to photocurrent generation. Therefore, when maximizing conversion efficiency, it is crucial that the low-energy cutoff be as close to, but still above the bandgap at the equilibrium cell temperature. While the general trend of conversion efficiency as a function of the high-energy cutoff can be seen in the lower panel of Fig. 4(b), this trend does not fully represent the severity of the sensitivity of $\eta$ to $E_{hc}$. In Fig. 4(b), the low-energy cutoff is always larger than the cell absorption bandgap. In reality, however, $E_g$ is sensitive to $E_{hc}$ as thermal losses, and thus the cell temperature, decrease as the high-energy cutoff decreases. For instance, if $E_{hc}$ is 0.01 eV and 0.02 eV lower than the optimal value, $E_g$ becomes larger than $E_{lc}$, such that the conversion efficiency reduces to 37.4% and 34.7%, respectively.

One must keep in mind that the optimal emission spectrum that includes all losses is specific to the case being studied in which $T_{rad}$, $h_\infty$ and $T_\infty$ have values of 2000 K, 600 Wm$^{-2}$K$^{-1}$ and 293 K, respectively. If $h_\infty$ is decreased while $T_{rad}$ and $T_\infty$ are held constant, the width of the optimal emission spectrum decreases due to an increasing $T_{cell}$ causing the dark current density to play a larger role at lower $E_{hc}$ values. For example, when $h_\infty$ is 100 Wm$^{-2}$K$^{-1}$, the optimal emission spectrum has $E_{lc}$ and $E_{hc}$ values of 0.721 eV and 0.731 eV resulting in a conversion efficiency of 34.8%. If $h_\infty$ is increased beyond 600 Wm$^{-2}$K$^{-1}$, the width of the optimal emission spectrum increases for the same reasoning. When $h_\infty$ is 1000 Wm$^{-2}$K$^{-1}$, the optimal emission spectrum has $E_{lc}$ and $E_{hc}$ values of 0.719 eV and 0.792 eV, and this results in a conversion efficiency of 39.4%. From this analysis, it is clear that as $h_\infty$ increases, the maximum possible conversion efficiency increases. However, as $h_\infty$ approaches infinity, $\eta$ converges to the value for the RE case of 40.8%.

## 3.2. Maximization of TPV output power density

The thermal emission spectrum maximizing output power density $P_m$ when radiative, electrical and thermal losses are accounted for is shown in Fig. 5. This emission spectrum has been obtained using 400 bands, instead of 125 bands, since the spectral range to be investigated for this case was determined to be 0.592 eV to 1.16 eV. Each spectral band has the same bandwidth as the case for maximizing $\eta$ of 1.42×10$^{-3}$ eV. Further refinement of the discretization by a factor of two resulted in a variation of the power density, the conversion efficiency and the width and spectral cutoffs of the optimal emission spectrum of less than 1%.

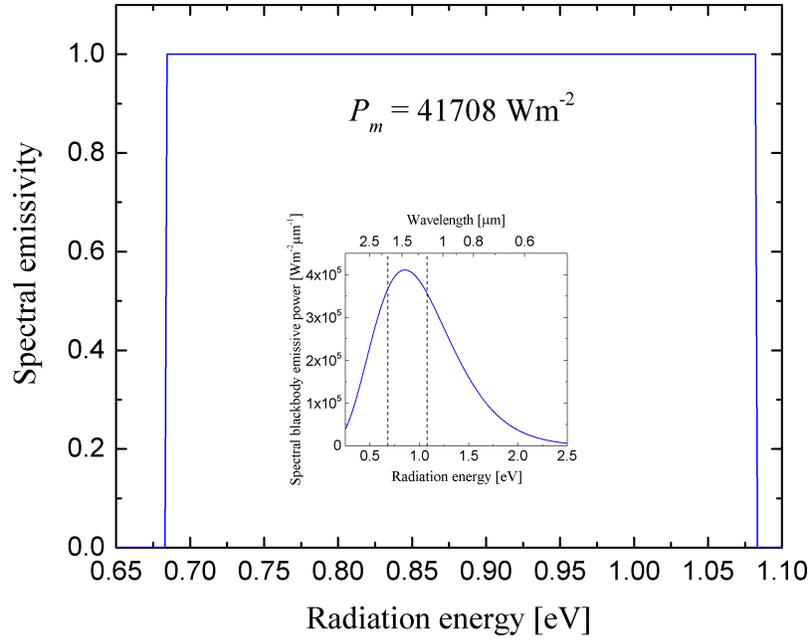

Figure 5. Thermal emission spectrum maximizing power density $P_m$ when radiative, electrical and thermal losses (RET) are considered. The inset shows the portion of the blackbody emissive power at 2000 K included in the optimal spectrum.

The emission spectrum maximizing $P_m$ takes the form of a step function where the emissivity of the radiator below 0.684 eV and above 1.082 eV is zero, while the emissivity between these limits has the maximum value of unity. The resulting power density, conversion efficiency and cell temperature are respectively 41708 Wm$^{-2}$, 24.4% and 393 K. Note that the low-energy cutoff is equal to the absorption bandgap of the GaSb cell at a temperature of 393 K. Here, the cell temperature is considerably larger than for the efficiency maximization case. This is due to the fact that the high-energy cutoff $E_{hc}$ is at a larger radiation energy, thus leading to significant thermal losses by thermalization and EHP recombination. The high-energy cutoff $E_{hc}$ of 1.082 eV for the optimal emission spectrum can be explained by analyzing Fig. 6, where the power density $P_m$, the current density at maximum power $J_m$, the voltage at maximum power $V_m$, and the dark current density $J_0(V_m)$ are plotted as a function of $E_{hc}$. For clarity, each curve is normalized by its own maximum.

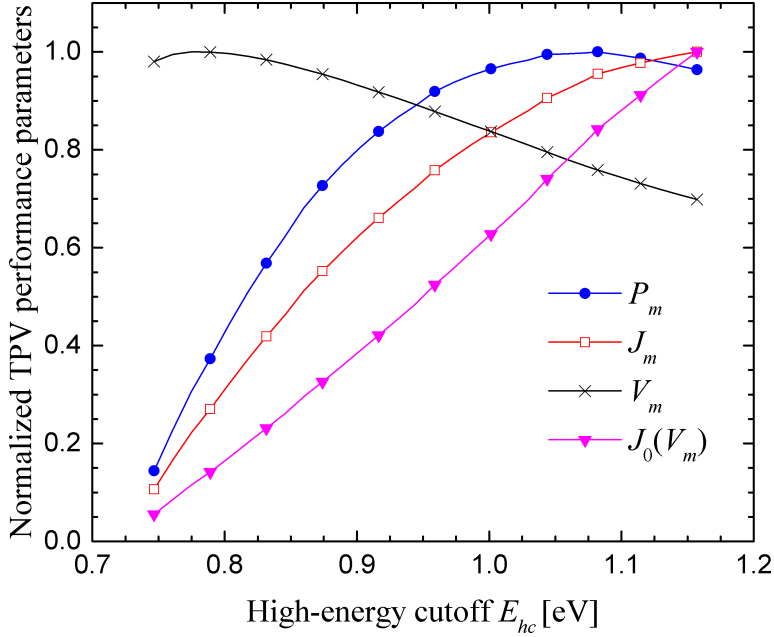

Figure 6. Power density $P_m$, current density at maximum power $J_m$, voltage at maximum power $V_m$, and dark current density $J_0(V_m)$ as a function of the high-energy cutoff $E_{hc}$. Each curve is normalized by its own maximum.

It can be seen in Fig. 6 that $J_m$ increases with $E_{hc}$ due to an increasing amount of radiation absorbed with energy larger than the cell bandgap, thus leading to a larger photocurrent generation. Increasing $E_{hc}$, however, has the drawback of raising the cell temperature due to increasing thermal losses by bulk and surface non-radiative recombination of EHPs and by thermalization, as shown in Fig. 7. The increasing cell temperature results in an increase of $J_0(V_m)$ which, in turn, causes $V_m$ to drop. Since the power density is the product of $J_m$ and $V_m$, there is a high-energy limit beyond which the decrease in $V_m$ outweighs the increase in $J_m$. This high-energy limit occurs at 1.082 eV when the decreasing normalized slope of the voltage $V_m$ is larger than the increasing normalized slope of the current density $J_m$ (i.e., when $(1/V_m)|dV_m/dE_{hc}| > (1/J_m)|dJ_m/dE_{hc}|$). Beyond this limit, absorption of radiation with energy larger than 1.082 eV is more detrimental than beneficial to the power density $P_m$ due to thermal losses.

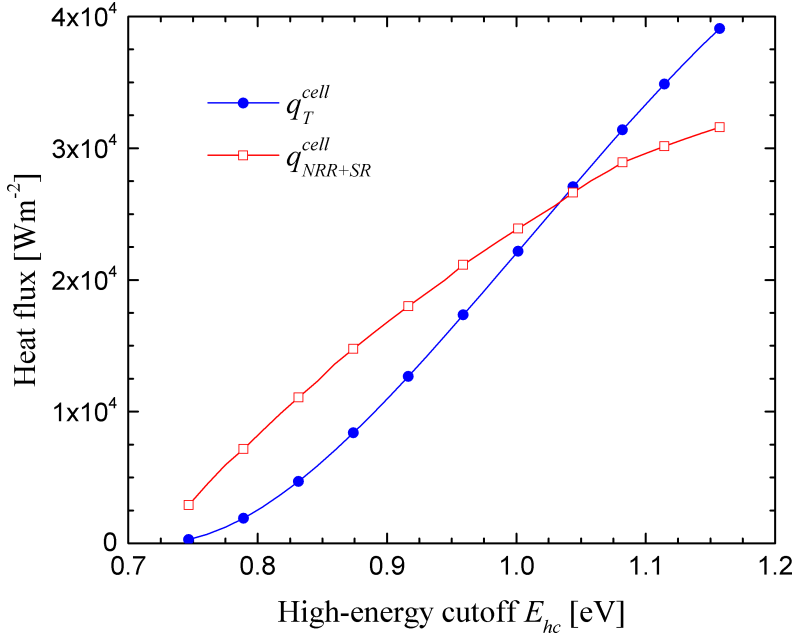

Figure 7. Heat fluxes absorbed in the cell due to thermalization $q_T^{cell}$ and non-radiative recombination (bulk and surface) $q_{NRR+SR}^{cell}$ as a function of the high-energy cutoff $E_{hc}$.

The emission spectrum maximizing TPV power density exhibits low- and high-energy cutoffs only when thermal losses are considered. Indeed, when only radiative losses or radiative and electrical losses are taken into account, there is no high-energy cutoff due to the fact that for a fixed $T_{cell}$, the power density continuously increases with increasing absorption of radiation with energy larger than the cell bandgap. Radiation with energy below the cell bandgap has no effect, positive or negative, on TPV power density when thermal losses are not considered. Therefore, there is no need for a low-energy cutoff. These results thus clearly demonstrate that it is imperative to account for thermal losses when designing TPV power generators maximizing output power density.

As the low-energy cutoff $E_{lc}$ decreases below the bandgap of the cell, the effect on power density is extremely small, since there is little absorption at energies directly beneath the bandgap. Therefore, while radiation absorption by the lattice and free carriers beneath the bandgap has no electrical benefit, it also has minimal negative impacts. For instance, the power density, originally of 41708 Wm$^{-2}$, only reduces to 41703 Wm$^{-2}$ and 41699 Wm$^{-2}$ as $E_{lc}$ is 0.01 eV and 0.02 eV below the bandgap. As $E_{lc}$ increases by 0.01 eV and 0.02 eV above the bandgap, the power density decreases more significantly to 40275 Wm$^{-2}$ and 38809 Wm$^{-2}$. The spectral region directly above the bandgap is where most of the radiation energy is used for EHP generation, thus causing the power density to significantly decrease if radiation in this region is removed. The sensitivity of $P_m$ to the high-energy cutoff can be seen in Fig. 6. If $E_{hc}$ is reduced by 0.01 eV and 0.02 eV below that of the optimal cutoff, $P_m$ reduces to 41618 Wm$^{-2}$ and 41504 Wm$^{-2}$. This decrease is not as large as that caused by increasing $E_{lc}$, but the photocurrent generated still outweighs the negative effects of thermal losses in this spectral range causing $P_m$ to decrease as $E_{hc}$ decreases. The power density reduces to 41555 Wm$^{-2}$ and 41401 Wm$^{-2}$ as $E_{hc}$ increases by 0.01 eV and 0.02 eV above the optimal value. In this spectral range, the negative effects of thermal losses outweigh the benefit of generating additional photocurrent. From this sensitivity analysis, it is clear that the conversion efficiency is more sensitive to the energy cutoffs of the optimal emission spectrum than the output power density.

### 3.3. Comparison of optimal emission spectra against blackbody and tungsten radiators

TPV conversion efficiency $\eta$ and power density $P_m$ obtained in sections 3.1 and 3.2 are shown in Fig. 8, and are compared against $\eta$ and $P_m$ values obtained with tungsten and blackbody radiators. In all cases, radiative, electrical and thermal losses are taken into account.

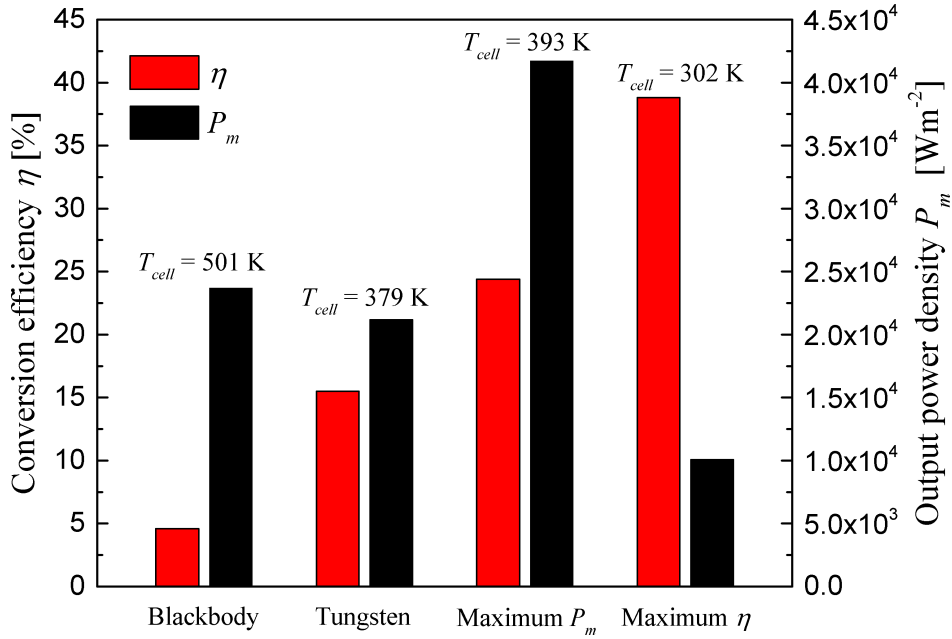

Figure 8. Power density $P_m$ and conversion efficiency $\eta$ obtained with various radiators: blackbody, tungsten, emission spectrum maximizing $P_m$ and emission spectrum maximizing $\eta$. In all cases, radiative, electrical and thermal losses (RET) are taken into account.

The emission spectrum maximizing $P_m$ leads to an output power density that is nearly twice that of the blackbody and tungsten radiators, and exceeds the output power density obtained with the emission spectrum maximizing $\eta$ by a factor of four. In addition, the cell temperature of 501 K with a blackbody radiator greatly exceeds that of the tungsten radiator (379 K), and the emission spectra maximizing $P_m$ (393 K) and $\eta$ (302 K). For the blackbody radiator, thermal emission below the cell bandgap, in addition to high thermal losses, result in a very low conversion efficiency of 4.6% compared to that of tungsten (15.5%) and the emission spectra maximizing $P_m$ (24.4%) and $\eta$ (38.8%).

The spectrum maximizing output power density $P_m$ has the benefit of operating with the second highest conversion efficiency. On the other hand, the spectrum maximizing conversion efficiency $\eta$ has the lowest $P_m$ of the radiators shown in Fig. 8, producing only less than half the $P_m$ of the tungsten radiator. However, an additional benefit of using the spectrum maximizing $\eta$ when operating a TPV power generator compared to the other radiators is that the cell temperature is very low. This will likely lead to a longer lifetime of the cell [29] in addition to a significant reduction to the cooling load. Ultimately, the choice between maximizing power density or conversion efficiency should be driven by the application. For waste heat recovery, a TPV device maximizing power density is the sensible choice, while maximization of the conversion efficiency should be considered for large-scale production of electricity via solar TPV systems.

## 4. Conclusions

The conversion efficiency and output power density of a thermophotovoltaic (TPV) device consisting of a spectrally selective radiator at a temperature of 2000 K, a gallium antimonide (GaSb) cell, and a cell thermal management system with a fluid temperature of 293 K and a heat transfer coefficient of 600 $Wm^{-2}K^{-1}$ have been maximized. This was accomplished by determining optimal radiator emission spectra using a framework in which a TPV model, accounting for radiative, electrical and thermal losses in the cell, is coupled with a genetic algorithm (GA). The results revealed that the emission spectrum maximizing conversion efficiency is not simply monochromatic at the cell absorption bandgap, but is instead a narrow step function where the radiator emissivity is unity between the absorption bandgap (0.719 eV at 302 K) and 0.763 eV, and zero outside that spectral band. This optimal thermal spectrum leads to TPV conversion efficiency and output power density of 38.8% and 10101 $Wm^{-2}$, respectively. In addition, it was shown that the high-energy cutoff of the optimal emission spectrum is highly sensitive to thermal losses in the cell, as an increase in the cell temperature negatively impacts the voltage at maximum power. This analysis thus clearly demonstrated that maximization of TPV conversion efficiency must account for radiative, electrical and thermal losses in the cell.

Maximization of TPV output power density is only possible when thermal losses are taken into account. Indeed, the effect of an increasing cell temperature allows the determination of a high-energy cutoff of the radiator emission spectrum beyond which radiation absorption is more detrimental than beneficial to the output power density. It was determined that the emission spectrum maximizing the output power density is a step function where the radiator emissivity is unity between the absorption bandgap (0.684 eV at 393 K) and 1.082 eV, and zero outside that spectral band. The optimal emission spectrum leads to an output power density of 41708 Wm$^{-2}$ and a conversion efficiency of 24.4%. The results also demonstrated that the optimal emission spectra obtained from the TPV-GA framework largely outperform TPV systems with tungsten and blackbody radiators.

From a practical standpoint, it is very difficult to find naturally occurring materials that are capable of producing the optimal thermal spectra determined with the TPV-GA framework. Therefore, man-made structures with designed radiative properties, such as those proposed in Refs. [4,7,10,14-21], are required. Promising work by Molesky et al. [20] and Sakr et al. [21] introduced radiator designs that could potentially produce emission spectra similar to those discussed in this paper. Using a metamaterial structure consisting of titanium nanowires embedded in a silicon host medium, Molesky et al. [20] designed a narrowband radiator with high emissivity in a spectral band close to that of the emission spectrum maximizing output power density. However, this specific design could not be operated at 2000 K, as it would exceed the melting temperature of silicon [30]. Sakr et al. [21] proposed a structure made of a rare-earth erbium-doped aluminum garnet (ErAG) wafer with a chirped, partially transmissive mirror on the emitting side and a highly-reflective dielectric mirror on the back side, resulting in a narrowband emission spectrum similar to the one maximizing conversion efficiency.

Alternatively, the design of radiator structures with thermal emission maximizing TPV output power density and conversion efficiency could likely be accomplished using a GA [31]. This is left as a future research effort.

**Acknowledgements**

This work was sponsored by the US Army Research Office under Grant no. W911NF-14-1-0210.